# A Função Relativística de Distribuição de Velocidades de Maxwell-Jüttner

Gilberto Jonas Damião, Clóves Gonçalves Rodrigues*

Pontifícia Universidade Católica de Goiás, CP 86, 74605-010, Goiânia, Goiás, Brazil

**Resumo**

Neste trabalho foi deduzida de maneira detalhada e didática a expressão da distribuição de velocidades relativística de Maxwell-Jüttner. Esta distribuição foi comparada com a bem conhecida distribuição de Maxwell-Boltzmann, a qual não leva em conta efeitos relativísticos. Aplicações e comparações entre as duas distribuições foram realizadas em três situações: *i*) prótons a temperatura da superfície solar, *ii*) elétrons à temperatura do critério de Lawson, e *iii*) prótons à temperaturas de quasares. Verificou-se que o valor da razão entre a massa de repouso da partícula considerada e a temperatura do sistema determina a necessidade de se considerar ou não um tratamento relativístico para a função de distribuição de velocidades do sistema.

Palavras-chave: Maxwell-Boltzmann, Maxwell-Jüttner, distribuição relativística, distribuição de velocidades.

# The Relativistic Maxwell-Jüttner Velocity Distribution Function

**Abstract**

In this paper, the expression of the relativistic Maxwell-Jüttner velocity distribution was deduced in a detailed and didactic way. This distribution was compared with the well-known Maxwell-Boltzmann distribution, which does not take into account relativistic effects. Applications and comparisons between the two distributions were carried out in three situations: *i*) protons at solar surface temperature, *ii*) electrons at Lawson's criterion temperature, and *iii*) protons at quasar temperatures. It was found that the value of the ratio between the rest mass of the particle considered and the temperature of the system determines the need to consider or not a relativistic treatment for the velocity distribution function of the system.

Keywords: Maxwell-Boltzmann, Maxwell-Jüttner, relativistic distribution, velocity distribution function.

* E-mail: cloves@pucgoias.edu.br

# 1. Introdução

A temperatura de qualquer sistema físico é o resultado do movimento de átomos e moléculas que compõem o sistema. Essas pequenas partes da matéria possuem diferentes velocidades, e a velocidade de cada partícula varia constantemente devido às colisões umas com as outras. Uma função de distribuição relativa à velocidade especifica a fração para cada intervalo de velocidades como função da temperatura do sistema. No início da segunda metade do século XIX, por volta de 1859, o físico e matemático britânico James Clerk Maxwell (1831-1879) realizou estudos sobre como se distribuíam os módulos das velocidades das moléculas de um gás em equilíbrio térmico e, no ano de 1860, demonstrou e divulgou que as velocidades das moléculas de um gás são distribuídas segundo a lei das distribuições dos erros, que foi formulada no ano de 1795 pelo matemático, físico e astrônomo alemão Johann Carl Friedrich Gauss (1777-1855). Nesta lei a energia cinética das moléculas é proporcional à temperatura absoluta $T$ do gás. Posteriormente, no ano de 1872, o físico austríaco Ludwig Eduard Boltzmann (1844-1906) generalizou esta lei, sendo atualmente conhecida como "lei de distribuição de velocidades de Maxwell-Boltzmann".

A distribuição de velocidades de Maxwell-Boltzmann é em suma uma distribuição de probabilidade que pode ser aplicada em diversas áreas da ciência. Esta função de distribuição de velocidades, preconizada por Maxwell e por Boltzmann, designada por $f_{MB}$, é dada por [1,2]

$$f_{MB}(v) = 4\pi \left(\frac{m}{2\pi kT}\right)^{3/2} v^2 e^{-\frac{mv^2}{2kT}}. \tag{1}$$

onde $v$ é a velocidade, $k$ é a constante de Boltzmann, $m$ é a massa da partícula e $T$ é a temperatura. O gráfico desta distribuição assemelha-se a uma gaussiana com início na origem, crescimento quadrático e decrescimento exponencial, como ilustrado na Fig. 1. Nota-se que o efeito do aumento da temperatura torna a curva de probabilidade mais achatada e longa. Fisicamente isto significa um aumento na probabilidade das partículas possuírem velocidades maiores com o aumento da temperatura.

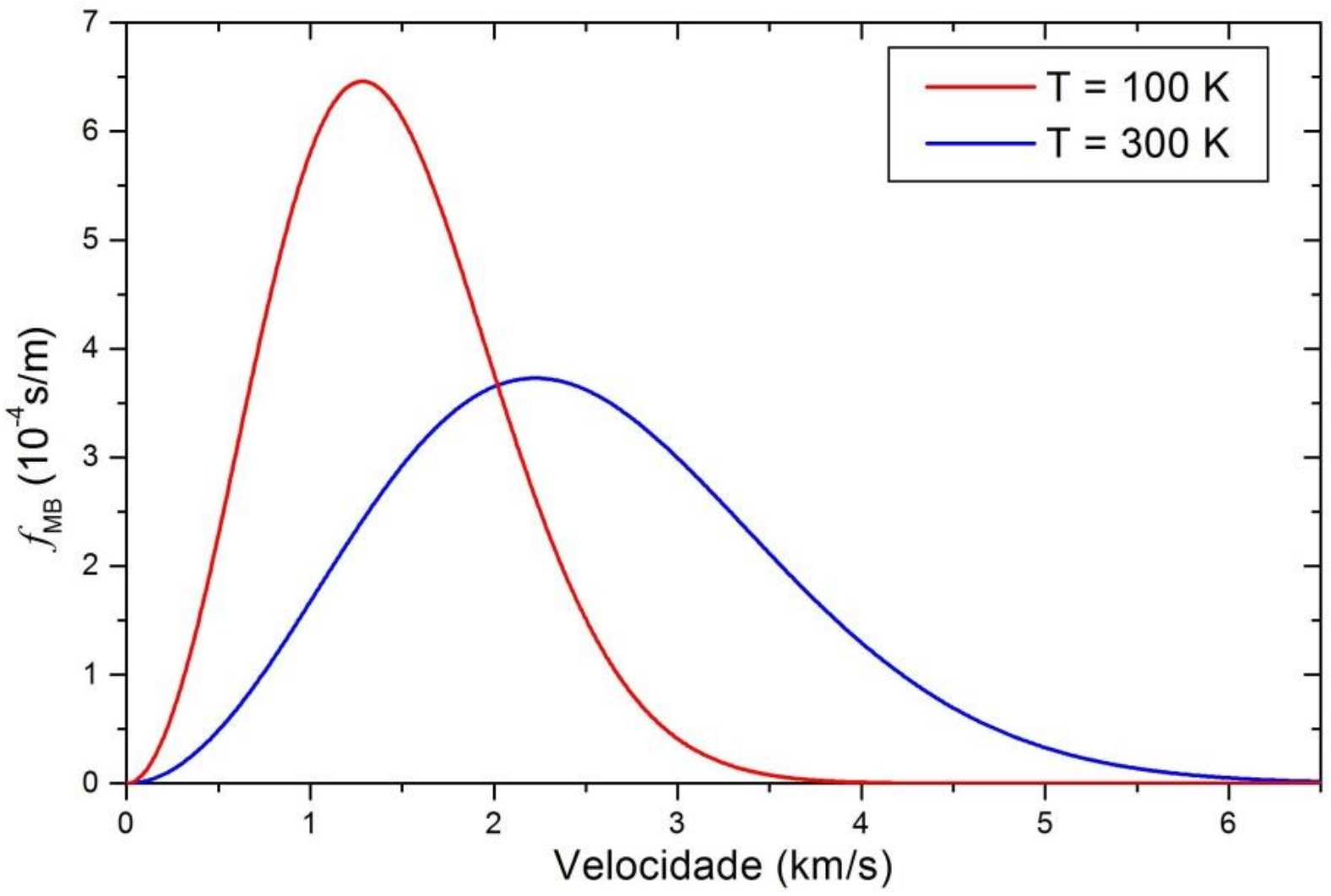

**Figura 1** *Função de distribuição de velocidades de Maxwell-Boltzmann para dois valores de temperatura. O valor adotado para a massa m foi a massa do próton.*

Uma maneira de verificar experimentalmente a validade da lei de distribuição de velocidades de Maxwell-Boltzmann consiste em analisar como variam alguns processos moleculares, como a velocidade das reações químicas quando a temperatura varia [3]. Outra maneira para uma verificação mais direta consiste na contagem do número de moléculas em cada intervalo de velocidades ou de energias. Na prática, isto pode ser feito com um método experimental que utiliza um seletor mecânico de velocidades composto por discos e fendas que giram com uma velocidade angular determinada, de tal forma que selecione as velocidades desejadas, por exemplo, para as partículas expelidas por um orifício de um tanque contendo um gás a uma temperatura $T$. Os valores experimentais obtidos nestes experimentos estão em excelente acordo com a teoria de Maxwell-Boltzmann [3].

Uma questão central é o fato da distribuição de velocidades de Maxwell-Boltzmann admitir uma probabilidade não nula de se encontrar partículas com velocidades maiores que a velocidade da luz no vácuo $c$. No entanto, sabe-se pela teoria da relatividade que somente velocidades menores que $c$ possuem sentido físico. Portanto, analisaremos neste artigo uma função de distribuição de velocidades capaz de lidar com este problema: a função de distribuição proposta por Ferencz Jüttner [4,5].

Esta função de distribuição de velocidades foi proposta por Jüttner em 1911, apenas seis anos após a formulação de Einstein da teoria da relatividade restrita [4]. Ela leva em consideração o limite da velocidade da luz, algo que a distribuição de velocidades de Maxwell-Boltzmann não faz. É possível deduzi-la de diferentes maneiras [5]. Entre estas, escolheu-se aqui, o tratamento físico-matemático mais acessível: a substituição da energia cinética clássica pela energia relativística na expressão do fator de Boltzmann e a consequente normalização desta função.

Existe uma grande quantidade de material bibliográfico relativo à distribuição de velocidades de Maxwell-Boltzmann. No entanto, com relação à distribuição de velocidades de Maxwell-Jüttner existem poucas publicações que a enfoquem da maneira que se optou por abordá-la neste trabalho. Quanto às suas aplicações, encontra-se na literatura aplicações da distribuição de velocidades de Maxwell-Jüttner em gases relativísticos [6], em rotações rígidas [7], em física de plasmas [8], dinâmica molecular em gases relativísticos [9,10], efeitos astrofísicos e de alta energia [11,12].

Este trabalho está assim organizado. Na Seção 2 a expressão da distribuição de velocidades relativística de Maxwell-Jüttner é deduzida de maneira detalhada e didática. Na Seção 3 são determinadas as expressões para os valores esperados de velocidade para esta distribuição. Na Seção 4 são feitas aplicações e comparações da distribuição de velocidades de Maxwell-Jüttner com a distribuição de velocidades de Maxwell-Boltzmann. A Seção 5 se reserva aos comentários finais.

## 2. A Função de Distribuição de Maxwell-Jüttner

O fator de Boltzmann, dado por $e^{-E/kT}$, compõe a expressão de distribuição de probabilidade Boltzmann [13], também conhecida como distribuição canônica ou distribuição de Gibbs [14], da seguinte forma

$$\Phi(p) = \frac{e^{-E/kT}}{Z}, \tag{2}$$

onde $Z$ é um fator de normalização.

A probabilidade $\Phi$ depende do momento linear $p$ das partículas. Tomando por $f_{MJ}(p)$ a função distribuição de momentos de Maxwell-Jüttner, tem-se, no espaço tridimensional de momentos, o seguinte fundamento probabilístico

$$f_{MJ}(p)dp = \Phi(p_x, p_{y,} p_z)dp_x dp_y dp_z,$$

$$f_{MJ}(p)dp = \Phi(p)p^2 d\Omega dp\,,$$

$$f_{MJ}(p)dp = 4\pi\Phi(p)p^2 dp\,. \tag{3}$$

Inserindo a Eq. (2) na Eq. (3) tem-se

$$f_{MJ}(p)dp = \frac{4\pi}{Z} e^{-E/kT} p^2 dp \,. \tag{4}$$

Calcula-se a constante de normalização $Z$ a partir da Eq. (4) da seguinte forma

$$\int_0^{+\infty} f_{MJ}(p)dp = 1,$$

$$\int_0^{+\infty} \frac{4\pi}{Z} e^{-E/kT} p^2 dp = 1$$

e isolando $Z$

$$Z = 4\pi \int_0^{+\infty} e^{-E/kT} p^2 dp. \tag{5}$$

A energia relativística, dada por $E = \gamma m_0 c^2$ [15,16], pode ser escrita em termos do momento linear da seguinte forma

$$E = m_0 c^2 \sqrt{1 + \frac{p^2}{{m_0}^2 c^2}} \,, \tag{6}$$

onde o fator de Lorentz, $\gamma$, em função do momento linear é

$$\gamma(p) = \sqrt{1 + \frac{p^2}{{m_0}^2 c^2}} \,. \tag{7}$$

Substituindo (6) em (5), tem-se a constante de normalização $Z$ em função do momento linear

$$Z = 4\pi \int_0^{+\infty} p^2 \exp\left\{-\frac{m_0 c^2}{kT} \sqrt{1 + \frac{p^2}{{m_0}^2 c^2}}\right\} dp. \tag{8}$$

Inserindo (6) e (8) em (4) tem-se

$$f_{MJ}(p)dp = \frac{p^2 \exp\left\{-\frac{m_0 c^2}{kT} \sqrt{1 + \frac{p^2}{{m_0}^2 c^2}}\right\}}{\int_0^{+\infty} p^2 \exp\left\{-\frac{m_0 c^2}{kT} \sqrt{1 + \frac{p^2}{{m_0}^2 c^2}}\right\} dp} dp \,. \tag{9}$$

Deve ser observado que se na expressão para a energia $E$ da Eq. (2), ao invés da energia total $\gamma m_0 c^2$, tivesse sido utilizada somente a expressão da energia cinética $m_0 c^2(\gamma - 1)$, o resultado obtido em (9) seria o mesmo, pois, o novo termo adicional $-m_0 c^2$ se cancelaria nas exponenciais que estão no numerador e denominador da Eq. (9).

Para determinar a distribuição de velocidades, há que se substituir a expressão do momento relativístico em função da velocidade, isto é, $p(v)$, bem como sua representação diferencial $dp$

$$p = \gamma m_0 v, \tag{10}$$

$$dp = m_0\gamma^3 dv, \tag{11}$$

onde

$$\gamma(v) = \frac{1}{\sqrt{1-\left(\frac{v}{c}\right)^2}}, \tag{12}$$

e

$$\gamma(\beta) = \frac{1}{\sqrt{1-\beta^2}}, \tag{13}$$

sendo

$$v = \beta c. \tag{14}$$

Para simplificar a notação $\gamma(v)$ e $\gamma(\beta)$ serão escritos apenas como $\gamma$. A substituição das Eqs. (10), (11) e (12) em (9) resulta em

$$f_{MJ}(v)dv = \frac{v^2\gamma^5 e^{\left(-\frac{m_0c^2}{kT}\gamma\right)}}{\int_0^c v^2\gamma^5 e^{\left(-\frac{m_0c^2}{kT}\gamma\right)} dv} dv\,. \tag{15}$$

Note que o momento linear, dado por $p = \gamma m_0 v$, tende a infinito quando a velocidade tende a $c$, pois nessa situação o fator de Lorentz tende a infinito. Por este motivo, o limite superior de integração da constante de normalização em (15) é $c$ [17].

A manipulação das equações com a variável $\beta$ ao invés de $v$ é mais adequada, uma vez que quando a velocidade tende a $c$, $\beta$ tende a 1, facilitando a análise gráfica e a relação entre as velocidades características como percentuais da velocidade da luz. Dessa forma, pela Eq. (14)

$$dv = cd\beta\,. \tag{16}$$

Substituindo (14) e (16) em (15) tem-se

$$f_{MJ}(\beta)d\beta = \frac{\beta^2\gamma^5 e^{\left(-\frac{m_0c^2}{kT}\gamma\right)}}{Z_{MJ}} d\beta\,, \tag{17}$$

onde

$$Z_{MJ} = \int_0^1 \beta^2\gamma^5 e^{\left(-\frac{m_0c^2}{kT}\gamma\right)} d\beta. \tag{18}$$

A constante de normalização $Z_{MJ}$ que figura no denominador da Eq. (17) apresenta uma integral não trivial. A sua determinação encontra-se no Apêndice A. Assim a temos

$$Z_{MJ} = \frac{K_2(\zeta)}{\zeta}\,, \tag{19}$$

onde no numerador $K_2$ é função modificada de Bessel de segunda espécie e de segunda ordem [18,19], cujo parâmetro variável é $\zeta$, que, por sua vez, depende da temperatura e da massa das partículas, expresso por

$$\zeta = \frac{m_0c^2}{kT} \tag{22}$$

A análise do parâmetro zeta, $\zeta$, é útil para determinar a necessidade de tratamento relativístico [20].

Substituindo (19) em (17) tem-se a distribuição de velocidades de Maxwell-Jüttner em função de $\beta$

$$f_{MJ}(\beta)d\beta = \frac{\zeta\beta^2\gamma^5 e^{-\zeta\gamma}}{K_2(\zeta)}d\beta\,, \tag{21}$$

ou apenas

$$f_{MJ}(\beta) = \frac{\zeta\beta^2\gamma^5 e^{-\zeta\gamma}}{K_2(\zeta)}\ . \tag{22}$$

Observe-se que, para retornar à distribuição de Maxwell-Jüttner cuja variável é a velocidade $v$, basta substituir $\beta$ por $v/c$ e $d\beta$ por $dv/c$ na Eq. (21), o que resulta em

$$f_{MJ}(v)dv = \frac{\zeta v^2\gamma^5 e^{-\zeta\gamma}}{c^3K_2(\zeta)}dv\,, \tag{23}$$

ou apenas

$$f_{MJ}(v) = \frac{\zeta v^2\gamma^5 e^{-\zeta\gamma}}{c^3K_2(\zeta)}\,. \tag{24}$$

Entretanto, quando se está lidando com um tratamento relativístico, é mais conveniente utilizar a distribuição em função de $\beta$ ao invés de $v$. Valiosas informações estatísticas podem ser extraídas da Eq. (22) a partir de valores esperados, o que será feito na próxima seção. Verifica-se que, para $v \ll c$, a distribuição de velocidades de Maxwell-Jüttner, Eq. (22), se reduz à de Maxwell-Boltzmann (veja Apêndice C).

## 3. Valores Esperados

Os valores esperados e sua dedução para a distribuição de Maxwell-Boltzmann são encontrados com maior facilidade na literatura [21]. Para a distribuição de Maxwell-Jüttner, contudo, não há muito material disponível [5,22]. Nesta seção, com base em $f_{MJ}(\beta)$ e em considerações estatísticas, determinaremos soluções analíticas para o valor mais provável, o valor médio e o valor médio quadrático em relação à variável $\beta$.

### 3.1 Valor mais provável de $\boldsymbol{\beta}$: $\boldsymbol{\beta_P}$

O valor mais provável de $\beta$ é obtido a partir da Eq. (22). Derivando-a e igualando-a a zero, tem-se

$$\frac{\partial}{\partial\beta}f_{MJ}(\beta) = 0,$$

$$\frac{\partial}{\partial\beta}\left(\frac{\zeta\beta^2\gamma^5 e^{-\zeta\gamma}}{K_2(\zeta)}\right) = 0,$$

$$\frac{\zeta}{K_2(\zeta)}\frac{\partial}{\partial\beta}\left[\frac{\beta^2 e^{-\zeta/\sqrt{1-\beta^2}}}{(1-\beta^2)^{5/2}}\right] = 0,$$

$$\frac{\zeta}{K_2(\zeta)}\left(\frac{e^{-\zeta/\sqrt{1-\beta^2}}[\beta^2\left(1-\zeta\sqrt{1-\beta^2}\right)-3\beta^4+2]\beta}{(1-\beta^2)^{9/2}}\right) = 0\,,$$

$$e^{-\zeta/\sqrt{1-\beta^2}}\left[\beta^2\left(1-\zeta\sqrt{1-\beta^2}\right)-3\beta^4+2\right]\beta = 0\,. \tag{25}$$

Para $\beta = 0$, a igualdade em (25) é satisfeita. Se $\beta$ tende a 1, o termo exponencial tende a zero, validando novamente a expressão. Logo, a análise pode se restringir ao termo

$$\beta^2 \left(1 - \zeta\sqrt{1-\beta^2}\right) - 3\beta^4 + 2 = 0\,, \qquad (26)$$

que pode ser simplificada para (veja Apêndice D)

$$9\beta^6 + (\zeta^2 + 3)\beta^4 - 8\beta^2 - 4 = 0\,. \qquad (27)$$

Entre as raízes da Equação (27), há uma entre 0 e 1 cuja expressão é dada por [23]

$$\beta_p = \sqrt{\sigma_1 + \sigma_2 + \sigma_3 + \sigma_4}\,, \qquad (28)$$

sendo $\beta_p$ o valor mais provável de $\beta$, e

$$\sigma_1 = -\frac{\zeta^2}{27} + \frac{\zeta^4}{27\sqrt[3]{\Delta}}\,, \qquad (29)$$

$$\sigma_2 = \frac{2\zeta^2}{9\sqrt[3]{\Delta}}\,, \qquad (30)$$

$$\sigma_3 = \frac{\sqrt[3]{\Delta}}{27}\,, \qquad (31)$$

e

$$\sigma_4 = -\frac{1}{9} + \frac{25}{3\sqrt[3]{\Delta}}\,, \qquad (32)$$

onde

$$\Delta = -\zeta^6 - 9\zeta^4 - 351\zeta^2 + 54\sqrt{3}\sqrt{-\zeta^6 - 13\zeta^4 - 375\zeta^2} + 3375\,.$$

### 3.2 Valor médio de $\boldsymbol{\beta}$: $\langle\boldsymbol{\beta}\rangle$

O valor médio, $\langle\beta\rangle$, pode ser determinado a partir da seguinte integração

$$\langle\beta\rangle = \int_0^1 \beta f_{MJ}(\beta)d\beta. \qquad (33)$$

Inserindo a Eq. (22) na equação anterior

$$\langle\beta\rangle = \int_0^1 \frac{\zeta\beta^3\gamma^5 e^{-\zeta\gamma}}{K_2(\zeta)} d\beta,$$

$$\langle\beta\rangle = \frac{\zeta}{K_2(\zeta)}\int_0^1 \beta^3\gamma^5 e^{-\zeta\gamma} d\beta,$$

$$\langle\beta\rangle = \frac{\zeta}{K_2(\zeta)}\int_0^1 \frac{\beta^3 e^{-\zeta/\sqrt{1-\beta^2}}}{(1-\beta^2)^{5/2}} d\beta, \qquad (34)$$

que resulta em (veja Apêndice E)

$$\langle\beta\rangle = \frac{\zeta}{K_2(\zeta)}\left(\frac{e^{-\zeta/\sqrt{1-\beta^2}}}{\zeta} - \frac{\Gamma\left(3,\zeta/\sqrt{1-\beta^2}\right)}{\zeta^3}\right)\Bigg|_0^1, \tag{35}$$

onde $\Gamma(a,b)$ designa a função gama incompleta [24]. Aplicando os limites de integração em (35), tem-se

$$\langle\beta\rangle = \frac{\zeta}{K_2(\zeta)}\left(\frac{\Gamma(3,\zeta)}{\zeta^3} - \frac{e^{-\zeta}}{\zeta}\right). \tag{36}$$

### 3.3 Valor da raiz quadrada do valor quadrático médio $\boldsymbol{\beta}$: $\boldsymbol{\beta_{rms}}$

Passemos para o cálculo de $\langle\beta^2\rangle$, que pode ser obtido pela expressão

$$\langle\beta^2\rangle = \int_0^1 \beta^2 f_{MJ}(\beta)d\beta.$$

Inserindo a Eq. (22) na equação anterior tem-se

$$\langle\beta^2\rangle = \int_0^1 \frac{\zeta\beta^4\gamma^5 e^{-\zeta\gamma}d\beta}{K_2(\zeta)}. \tag{37}$$

É possível expressar $\langle\beta^2\rangle$ a partir de funções matemáticas especiais que, em última análise, são séries de potências que entregam resultados a partir de resoluções numéricas. Utilizando a resolução exposta no Apêndice F, tem-se

$$\langle\beta^2\rangle = 1 - \frac{\zeta K_1(\zeta)}{K_2(\zeta)} + \frac{\zeta\pi}{2K_2(\zeta)}[1 - \zeta K_0(\zeta)L_{-1}(\zeta) - \zeta K_1(\zeta)L_0(\zeta)], \tag{38}$$

onde $L_\nu(\zeta)$ são funções modificadas de Struve [25] (veja Apêndice G). Definindo $\beta_{rms}$ (em analogia com $v_{rms}$) como

$$\beta_{rms} = \sqrt{\langle\beta^2\rangle}, \tag{39}$$

tem-se

$$\beta_{rms} = \sqrt{1 - \frac{\zeta K_1(\zeta)}{K_2(\zeta)} + \frac{\zeta\pi}{2K_2(\zeta)}[1 - \zeta K_0(\zeta)L_{-1}(\zeta) - \zeta K_1(\zeta)L_0(\zeta)]}. \tag{40}$$

## 4. Aplicações e Comparações entre as Distribuições de Velocidades de Maxwell-Jüttner e Maxwell-Boltzmann

### 4.1 Comparativo entre as distribuições de Maxwell-Jüttner e Maxwell-Boltzmann

As distribuições de velocidades apresentadas são utilizadas em situações específicas. Os fatores que determinam o tratamento clássico ou relativístico são a massa das partículas e a temperatura do sistema. O parâmetro zeta, $\zeta = m_0c^2/kT$, apresentado na Eq. (20) contém tais variáveis de interesse. Note que $\zeta$ é a razão entre a energia de repouso e a energia térmica do siste ma, e para corpúsculos de mesma massa, a temperatura absoluta é o fator determinante do regime do sistema [26]. A análise desta medida é essencial para determinar se um sistema será tratado de forma clássica ou relativística. Quando $\zeta$ tende a valores muito grandes, o tratamento não relativístico é o indicado. Regimes ultra-relativísticos ocorrem quando $\zeta$ tende a zero [20].

Para efeito de comparação, é mais conveniente que as distribuições de velocidades de Maxwell-Boltzmann e de Maxwell-Jüttner estejam em função das mesmas variáveis. De acordo com a dedução apresentada no Apêndice H, a distribuição de velocidades de Maxwell-Boltzmann em termos de $\beta$ e de $\zeta$ possui a seguinte forma

$$f_{MB}(\beta) = \sqrt{2/\pi}\zeta^{3/2}\beta^2 e^{-\zeta\beta^2/2}, \tag{41}$$

com os seguintes valores médios [21]

$$\beta_{p(MB)} = \sqrt{\frac{2kT}{m_0 c^2}} = \sqrt{\frac{2}{\zeta}},$$

$$\langle\beta\rangle_{MB} = \sqrt{\frac{8kT}{\pi m_0 c^2}} = \sqrt{\frac{8}{\pi\zeta}},$$

e

$$\beta_{rms(MB)} = \sqrt{\langle\beta^2\rangle_{MB}} = \sqrt{\frac{3kT}{m_0 c^2}} = \sqrt{\frac{3}{\zeta}}.$$

A Tabela 1 mostra um comparativo entre as distribuições de Maxwell-Boltzmann e Maxwell-Jüttner. As expressões apresentadas na Tabela 1 serão utilizadas na próxima seção para calcular os valores esperados de $\beta$ e para construir os gráficos das duas distribuições. Ressaltamos que as expressões obtidas para os valores médios usando a distribuição de velocidades de Maxwell-Boltzmann (segunda coluna da Tabela 1) podem também serem determinados diretamente a partir das expressões encontradas utilizando a distribuição de velocidades de Maxwell-Jüttner (terceira coluna da Tabela 1). Para tanto é necessário utilizar os limites das funções especiais $K_0(\zeta)$, $K_1(\zeta)$, $K_2(\zeta)$, $L_{-1}(\zeta)$, $L_0(\zeta)$, $\Gamma(3,\zeta)$ quando $v/c \ll 1$.

**Tabela 1** Comparativo entre as distribuições de Maxwell-Boltzmann e Maxwell-Jüttner.

| | Maxwell-Boltzmann | Maxwell-Jüttner |
|---|---|---|
| Distribuição | $\sqrt{2/\pi}\zeta^{3/2}\beta^2 e^{-\zeta\beta^2/2}$ | $\frac{\zeta\beta^2\gamma^5 e^{-\zeta\gamma}}{K_2(\zeta)}$ |
| $\beta_p$ | $\sqrt{\frac{2}{\zeta}}$ | $\sqrt{\sigma_1+\sigma_2+\sigma_3+\sigma_4}$ |
| $\langle\beta\rangle$ | $\sqrt{\frac{8}{\pi\zeta}}$ | $\frac{\zeta}{K_2(\zeta)}\left(\frac{\Gamma(3,\zeta)}{\zeta^3}-\frac{e^{-\zeta}}{\zeta}\right)$ |
| $\beta_{rms}$ | $\sqrt{\frac{3}{\zeta}}$ | $\sqrt{1-\frac{\zeta K_1(\zeta)}{K_2(\zeta)}+\frac{\zeta\pi}{2K_2(\zeta)}[1-\zeta K_0(\zeta)L_{-1}(\zeta)-\zeta K_1(\zeta)L_0(\zeta)]}$ |

A distribuição de Maxwell-Jüttner, dada pela Eq. (22), possui a seguinte característica: no limite de baixas velocidades ($v \ll c$ e, consequentemente, $v/c \ll 1$), $f_{MJ}(\beta)$ tende à distribuição de Maxwell-Boltzmann (veja Apêndice C), e nessa situação tem-se que $f_{MJ}(\beta) \approx f_{MB}(\beta)$, sendo que tal limite ocorre quando $\zeta \gg 0$ [20,26].

Analisaremos nas próximas seções, como aplicações, as distribuições para valores de temperatura que ocorrem na Astrofísica e na Física de Plasma de fenômenos de altas energias e temperaturas.

## 4.2 Temperatura da superfície do Sol

A temperatura da superfície do Sol é de cerca de $5.800$ K [27]. Para prótons (íons de hidrogênio), de massa individual de repouso $1{,}6726 \times 10^{-27}$ kg, tem-se que $\zeta \simeq 1{,}877 \times 10^{9}$, um valor muito grande. Trata-se, nesse caso, de um regime clássico, no qual as duas distribuições praticamente coincidem, como mostra a Fig. 2. Na Fig. 2 a curva em vermelho é a distribuição de velocidades de Maxwell-Jüttner, a qual coincide com a de Maxwell-Boltzmann, como pode ser visto pela curva em azul do gráfico menor, que mostra a razão $f_{\mathrm{MJ}}/f_{\mathrm{MB}} \simeq 1$. Os valores esperados de velocidade ($v_p$, $\langle v \rangle$ e $v_{rms}$) para as duas distribuições são mostrados na Tabela 2. Os valores esperados apresentaram-se praticamente idênticos em magnitude e em localização.

**Tabela 2** Valores esperados de velocidade (em m/s) de prótons para $T = 5.800$ K; ($\zeta \simeq 1{,}877 \times 10^{9}$).

| Valores esperados | Maxwell-Boltzmann | Maxwell-Jüttner | Diferença (%) |
|---|---|---|---|
| $v_p$ | 9.786,554 | 9.785,136 | 0,01 |
| $\langle v \rangle$ | 11.042,943 | 11.042,943 | 0 |
| $v_{rms}$ | 11.986,032 | 11.986,032 | 0 |

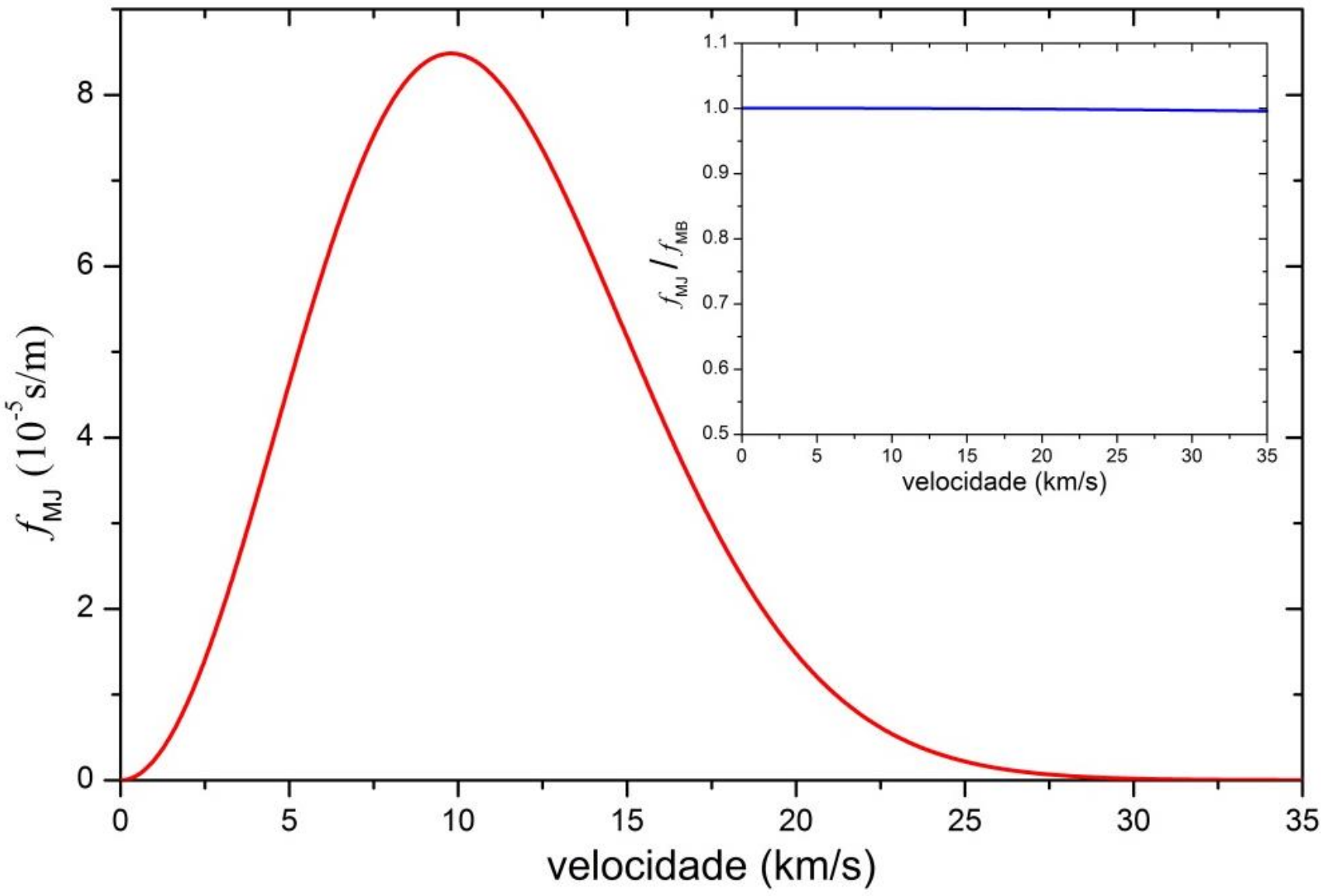

**Figura 2** Distribuição de velocidades de Maxwell-Jüttner para prótons ($m_0 = 1{,}6726 \times 10^{-27}$ kg) à temperatura da superfície do Sol ($T = 5.800$ K).

## 4.3 Temperatura de Reações D-D (deutério-deutério)

A próxima situação a ser analisada trata da temperatura ideal para ocorrência de fusão nuclear auto-sustentável. De acordo com Zohuri, a fusão nuclear controlada pode produzir mais energia que os processos de fissão nuclear [28]. O processo de fusão nuclear necessita de altas temperaturas para que o potencial elétrico de repulsão entre os prótons dos núcleos atômicos seja superado pela energia cinética destes. Trata-se de temperaturas capazes de ionizar completamente os átomos envolvidos, resultando em plasma de núcleos e elétrons. Para enfrentar menos repulsão e despender menos energia, utilizam-se elementos com poucos prótons no núcleo, como os isótopos do hidrogênio deutério e trítio. O deutério possui um próton e um nêutron em seu núcleo e o trítio um próton e dois nêutrons [29].

As reações D-D (deutério-deutério) e D-T (deutério-trítio) são bastante promissoras no que tange à relação entre o gasto energético para a ocorrência da fusão e a produção de energia neste processo. Considerando as variáveis temperatura, densidade e tempo de confinamento, o físico John D. Lawson publicou, em 1957, uma medida para que a energia produzida no processo de fusão superasse as perdas por radiação e mantivesse o aquecimento do plasma a fim de observar a auto-sustentabilidade da reação [27]. Tal medida denomina-se critério de Lawson [30], em que o produto entre a densidade $n$ do plasma e o tempo de confinamento $\tau$ daquele deve ser, para a reação D-T, $n\tau > 10^{20}\ \mathrm{s/m^3}$, com temperaturas da ordem de $10^7$ K e, para a reação D-D, $n\tau > 10^{22}\ \mathrm{s/m^3}$, considerando temperaturas da ordem de $10^8$ K.

Considerando um plasma de dêuterons e elétrons de reações D-D, a partir do critério de Lawson, estima-se que a temperatura mínima para que ocorra fusão nuclear é de $1{,}5 \times 10^8$ K [28]. A Tabela 3 apresenta o parâmetro $\zeta$ das partículas envolvidas. Observa-se que o parâmetro $\zeta$ para o dêuteron tem valor muito alto, o que indica um regime não-relativístico de velocidade das partículas.

**Tabela 3** Parâmetro $\zeta$ (reação D-D).

| | Massa (kg) | Temperatura (K) | $\zeta$ |
|---|---|---|---|
| Dêuteron | $3{,}3436 \times 10^{-27}$ | $1{,}5 \times 10^{8}$ | $1{,}45 \times 10^{5}$ |
| Elétron | $9{,}1094 \times 10^{-31}$ | $1{,}5 \times 10^{8}$ | 39,533 |

Já os elétrons, dependendo da temperatura do plasma, podem alcançar velocidades percentualmente consideráveis em relação a $c$. A Fig. 3 mostra os gráficos das distribuições de Maxwell-Boltzmann (em vermelho) e de Maxwell-Jüttner (em azul) em função de $\beta$ para elétrons nas condições da Tabela 3, isto é, para $T = 1{,}5 \times 10^8$. Nota-se que as duas distribuições têm o mesmo comportamento e para $\beta < 0{,}2$, as duas distribuições são praticamente idênticas. Os valores esperados de $\beta$ para as duas distribuições são mostrados na Fig. 3 e os seus valores na Tabela 4. Nota-se pela Tabela 4 que os valores esperados entre as duas distribuições possuem diferenças menores que 3%.

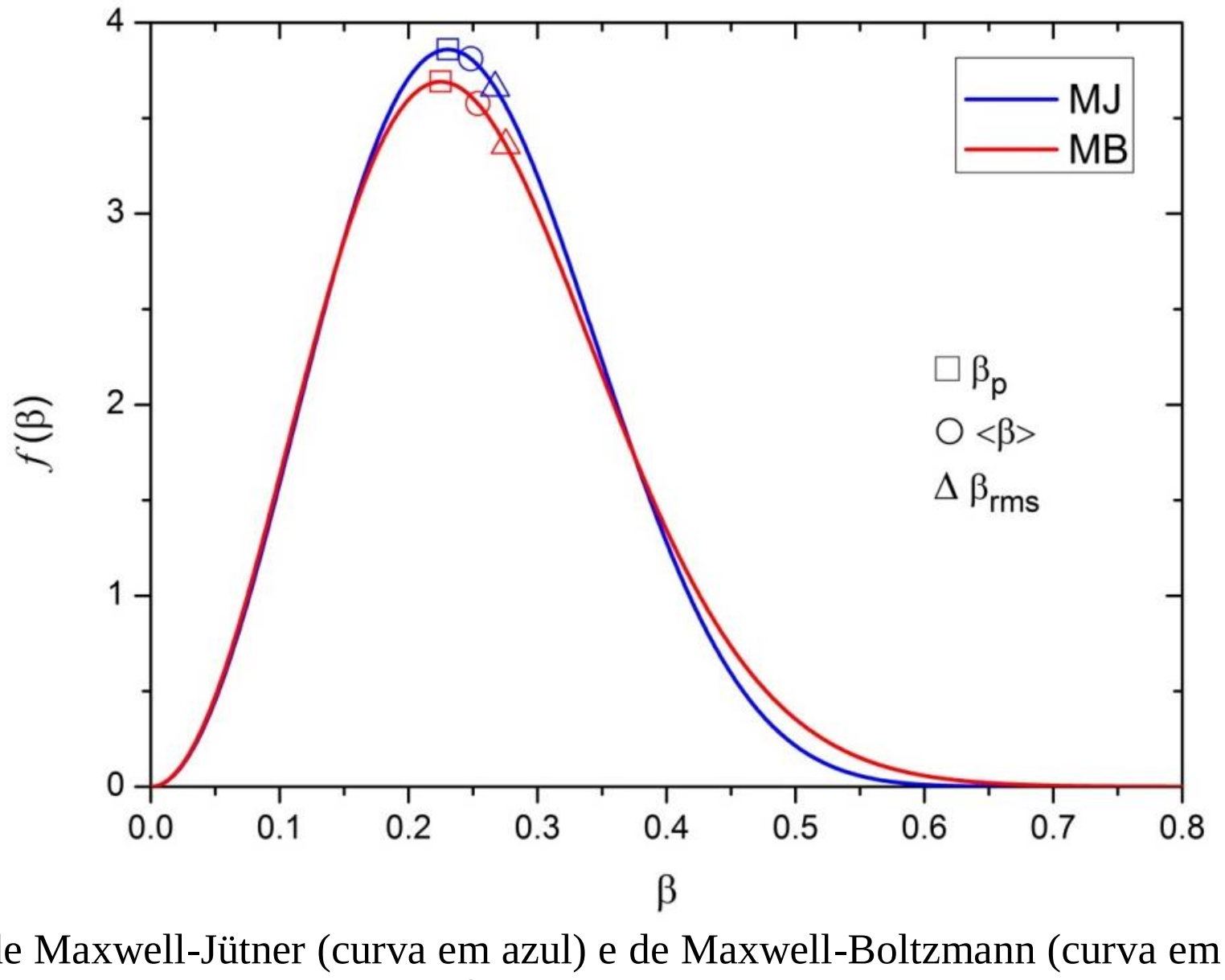

**Figura 3** Distribuições de Maxwell-Jütner (curva em azul) e de Maxwell-Boltzmann (curva em vermelho) para elétrons à temperatura do critério de Lawson ($T = 1{,}5 \times 10^8$ K); ($\zeta = 39{,}533$).

**Tabela 4** Valores esperados de $\beta$ para elétrons ($m_0 = 9{,}1094 \times 10^{-31}$ kg) com $T = 1{,}5 \times 10^8$ K; ($\zeta = 39{,}533$).

| Valores esperados | Maxwell-Boltzmann | Maxwell-Jüttner | Diferença (%) |
|---|---|---|---|
| $\beta_p$ | 0,2249 | 0,2305 | 2,4 |
| $\langle\beta\rangle$ | 0,2538 | 0,2483 | 2,2 |
| $\beta_{rms}$ | 0,2755 | 0,2672 | 3,0 |

O aumento da temperatura poderia fazer cessar a fusão nuclear auto-sustentada de D-D [31]. Temperaturas da ordem de $10^9$ K resultariam em perdas energéticas similares à energia produzida [30]. Os elétrons de uma reação D-D em uma temperatura de dez vezes o valor apresentado na Tabela 3, isto é, com $T = 1{,}5 \times 10^9$ K, apresentariam $\zeta$ dez vezes menor, ou seja, $\zeta = 3{,}9533$. Neste caso, a Fig. 4 mostra uma diferença maior entre as curvas das distribuições, e consequentemente os valores esperados de $\beta$ mostrados na Tabela 5 apresentam também uma diferença maior que na última situação analisada (vide Tabela 4). A distribuição de Maxwell-Jüttner, contudo, não mantém o mesmo formato para valores de $\zeta$ próximos de 1. Observam-se curvas com características e valores esperados de distribuições assimétricas à esquerda [32], com o valor da moda $\beta_p$ maior que o da média $\langle\beta\rangle$, e consequantemente $\langle\beta\rangle - \beta_p < 0$, como pode ser notado na Fig. 4.

Observa-se na Fig. 4 que a curva da distribuição de Maxwell-Boltzmann continua a ter um crescimento quadrático e um decrescimento exponencial, independe do valor de $\zeta$. A média $\langle\beta\rangle$ dessa distribuição é sempre maior que sua moda $\beta_p$, o que lhe atribui assimetria positiva ou à direita [32], pois $\langle\beta\rangle - \beta_p > 0$.

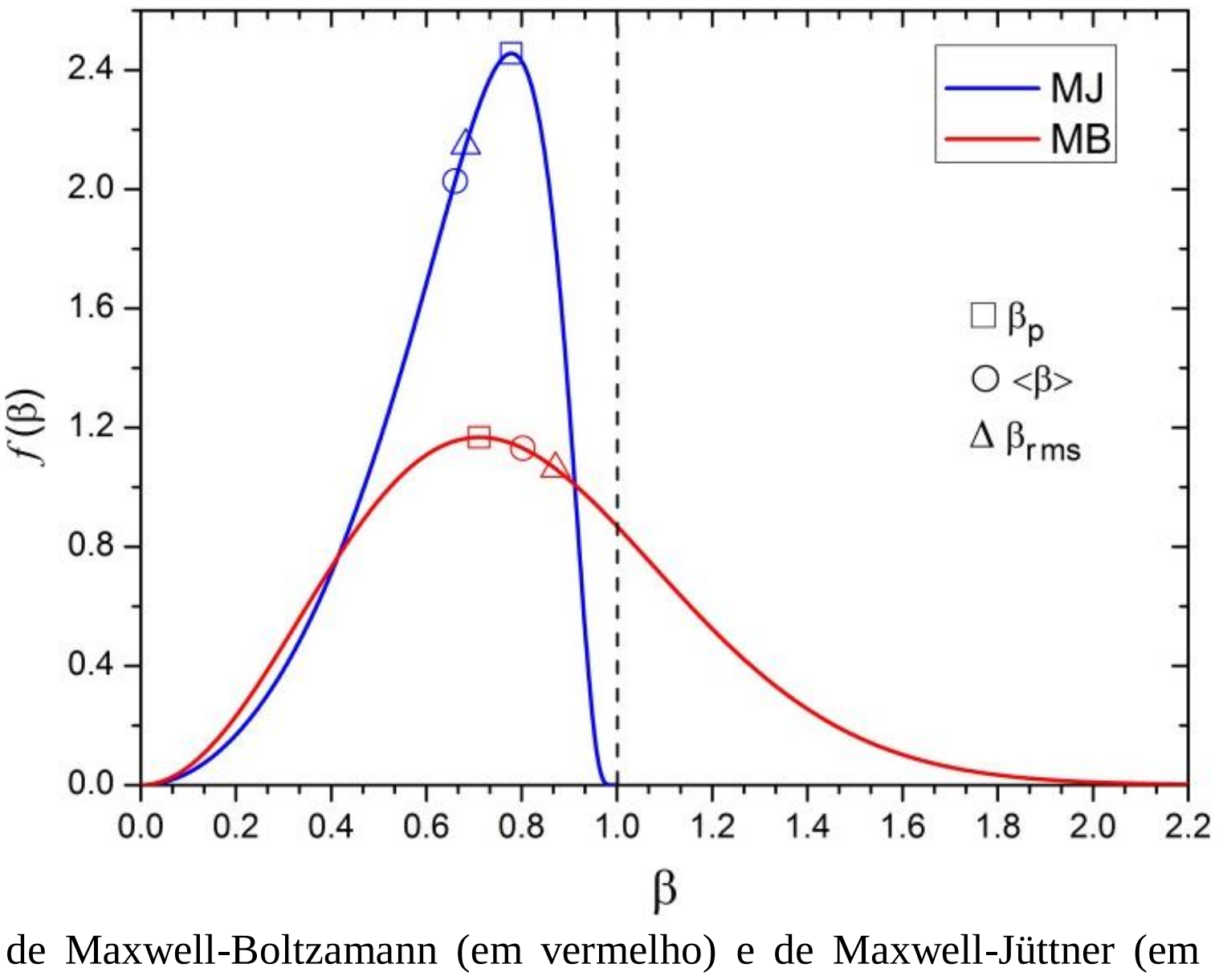

**Figura 4** Distribuições de Maxwell-Boltzamann (em vermelho) e de Maxwell-Jüttner (em azul) para elétrons com $T = 1{,}5 \times 10^9$ K, $\zeta = 3{,}9533$.

**Tabela 5** Valores esperados de $\beta$ para elétrons ($m_0 = 9{,}1094 \times 10^{-31}$ kg) com $T = 1{,}5 \times 10^9$ K; ($\zeta = 3{,}9533$).

| Valores esperados | Maxwell-Boltzmann | Maxwell-Jüttner | Diferença (%) |
|---|---|---|---|
| $\beta_p$ | 0,7113 | 0,7786 | 8,6 |
| $\langle\beta\rangle$ | 0,8027 | 0,6604 | 17,7 |
| $\beta_{rms}$ | 0,8712 | 0,6828 | 21,6 |

Ainda com respeito à distribuição de Maxwell-Boltzmann, os valores de $\beta$ maiores ou iguais a 1 não possuem sentido físico. Levando isso em consideração, o percentual de observações perdido com a utilização desta distribuição para $\zeta = 39{,}5335$, situação da Fig. 3, é de $1{,}3379 \times 10^{-8}$. Já para $\zeta$ dez vezes menor ($\zeta = 3{,}95335$), a perda é superior a um quarto, pois

$$\int_{1}^{+\infty} \sqrt{\frac{2}{\pi}}(3{,}9533)^{3/2}\beta^2 e^{-3{,}9533\beta^2/2}\, d\beta = 0{,}2665\,.$$

Tal perda fica mais clara quando se limita a abscissa $\beta$ a 1, como feito pela linha tracejada na Fig. 4 e pela área quadriculada sob a curva da Fig. 5 para a distribuição de Maxwell-Boltzmann para elétrons com $T = 1{,}5 \times 10^9$ K.

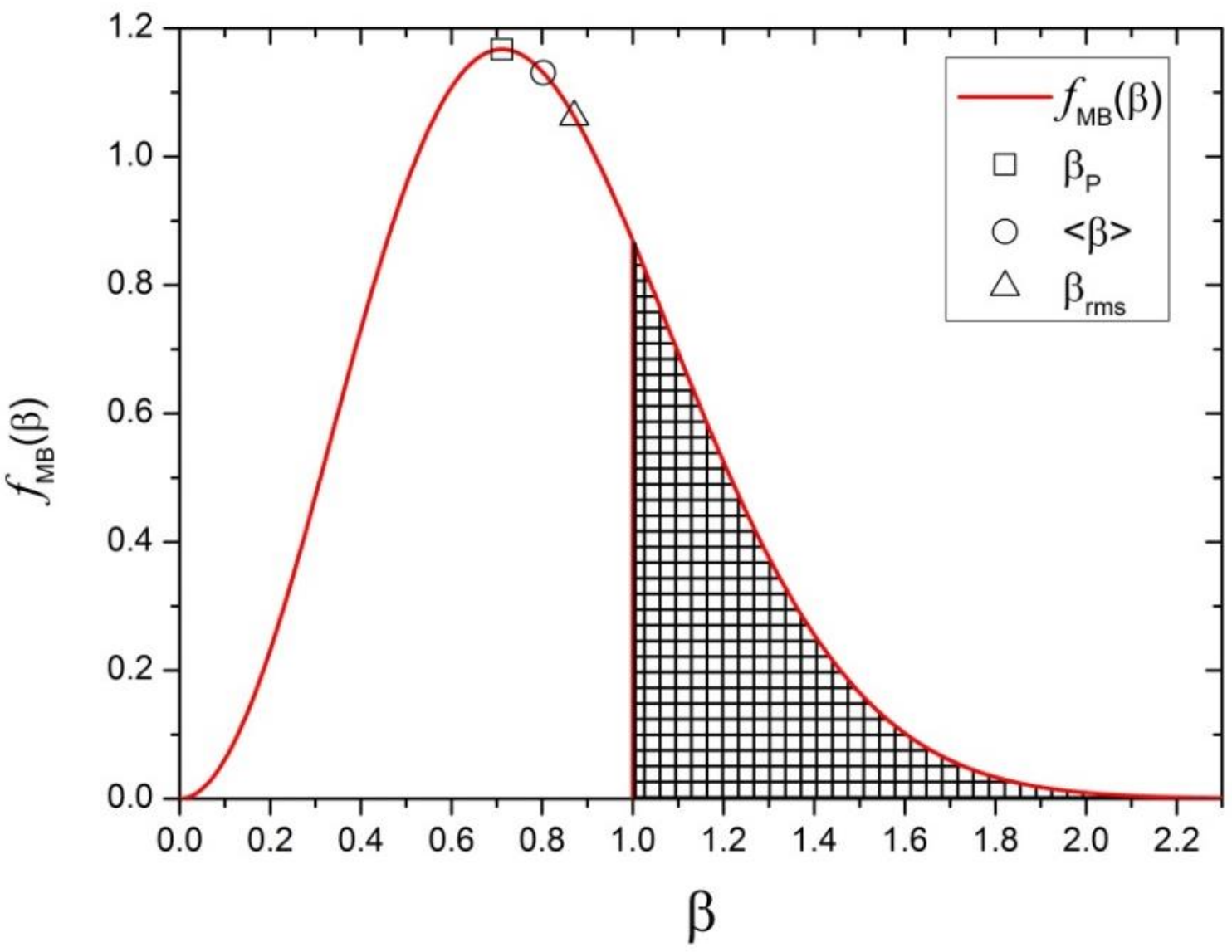

**Figura 4** Área perdida (quadriculada) pelo uso da distribuição de Maxwell-Boltzmann com $\zeta = 3{,}9533$, para elétrons ($m_0 = 9{,}1094 \times 10^{-31}$ kg) com $T = 1{,}5 \times 10^9$ K.

Pela Tabela 3, observa-se que, para uma mesma temperatura, o valor do parâmetro $\zeta$ varia somente com o valor da massa das partículas envolvidas. Partículas massivas como prótons precisariam de temperaturas bem maiores que as dos elétrons do plasma de fusão nuclear para que o valor de $\zeta$ se aproximasse de 1.

## 4.4 Temperatura em Quasares

Os quasares são objetos astrofísicos muito distantes da Terra e altamente luminosos [33]. Eles se localizam nos centros de algumas galáxias e são compostos por um buraco negro e uma imensa nuvem de gás que o orbita. A atração gravitacional do buraco negro faz com que o gás seja sugado e, nesse processo, as partículas do gás atingem altíssimas temperaturas e velocidades.

Entre os quasares conhecidos, o 3C 273 é o que possui maior luminosidade. Em razão disso, é um dos mais estudados. Há pesquisas que estimam que este objeto pode apresentar temperaturas da ordem de $10^{13}$ K [34]. Tal valor de temperatura pode provocar velocidades muito altas dos átomos e dos íons que compõem o gás no entorno do buraco negro. Prótons à temperatura $T = 10^{13}$ K possuem $\zeta = 1{,}0888$. A Fig. 6 mostra as distribuições de velocidades de Maxwell-Boltzmann (em vermelho) e de Maxwell-Jüttner (em azul) para tais parâmetros. A Tabela 6 mostra os valores obtidos para os valores médios de $\beta$ para as duas distribuições. Nota-se que, neste caso, com $\zeta = 1{,}0888$, existe uma grande diferença percentual entre os valores esperados das duas distribuições, chegando a 45% para $\beta_{rms}$.

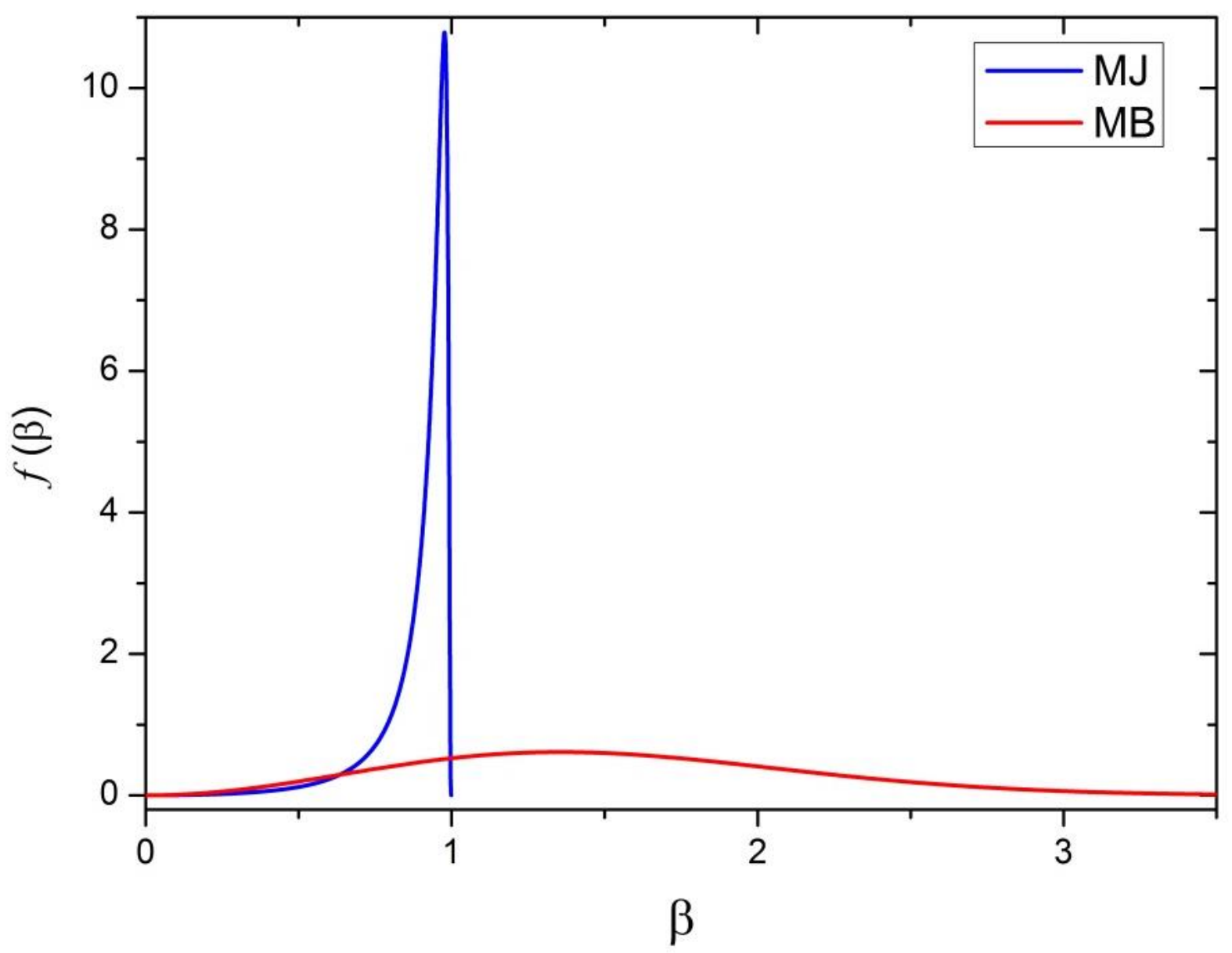

**Figura 6** Comparativo entre as distribuições de Maxwell-Boltzmann (em vermelho) e de Maxwell-Jüttner (em azul) para prótons ($m_0 = 1{,}6726 \times 10^{-27}$ kg) com $T = 10^{13}$ K; $\zeta = 1{,}0888$.

**Tabela 6** Valores esperados de $\beta$ para prótons com $T = 10^{13}$ K; ($\zeta = 1{,}0888$).

| Valores esperados | Maxwell-Boltzmann | Maxwell-Jüttner | Diferença (%) |
|---|---|---|---|
| $\beta_p$ | 1,3553 | 0,9769 | 27,9 |
| $\langle\beta\rangle$ | 1,5293 | 0,8953 | 41,5 |
| $\beta_{rms}$ | 1,6599 | 0,9024 | 45,6 |

Todos os valores esperados da distribuição de Maxwell-Boltzmann para prótons à temperatura do quasar 3C 273 não possuem sentido físico, pois o pico da curva, ocorre em $\beta > 1$, como pode ser melhor evidenciado na Fig. 7.

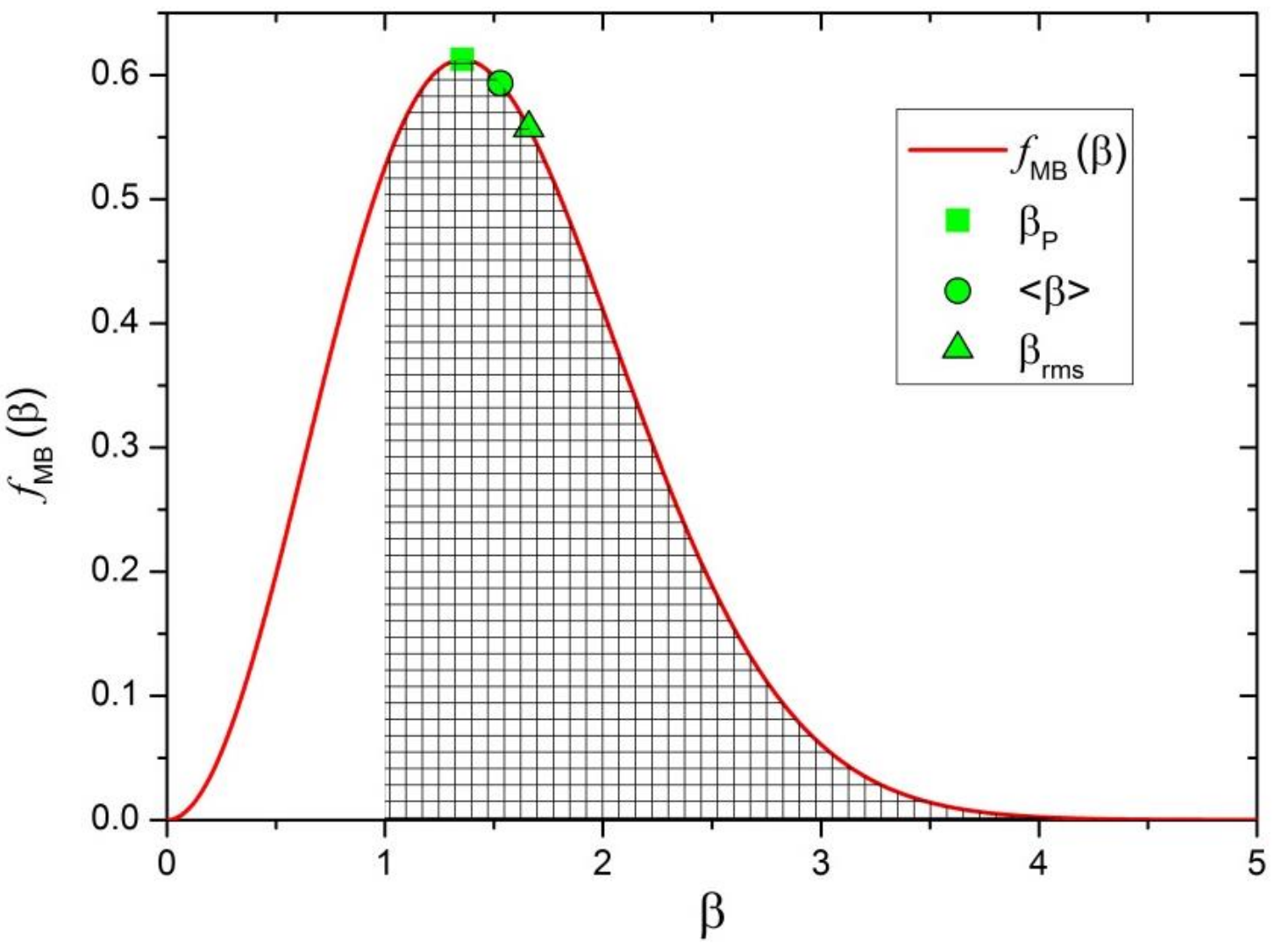

**Figura 7** Distribuição de Maxwell-Boltzmann para prótons ($m_0 = 1{,}6726 \times 10^{-27}$ kg) com $T = 10^{13}$ K; ($\zeta = 1{,}0888$).

As observações perdidas neste caso pela utilização da distribuição de Maxwell-Boltzmann com um parâmetro $\zeta$ tão pequeno, indicativo de um regime relativístico, seriam superiores a três quartos do total, pois

$$\int_{1}^{+\infty} \sqrt{\frac{2}{\pi}}(1{,}0888)^{3/2}\beta^2 e^{-1{,}0888\beta^2/2}\, d\beta = 0{,}7798\,.$$

Este percentual perdido é ilustrado pela área quadriculada na Fig. 7.

A distribuição de Maxwell-Jüttner de prótons à temperatura do quasar 3C 273 tem a curva e os valores esperados mostrados na Fig. 8. Comparando com os resultados obtidos nas Seções 4.2 e 4.3, nota-se claramente o deslocamento para a direita do pico da função quando o valor de $\zeta$ diminui.

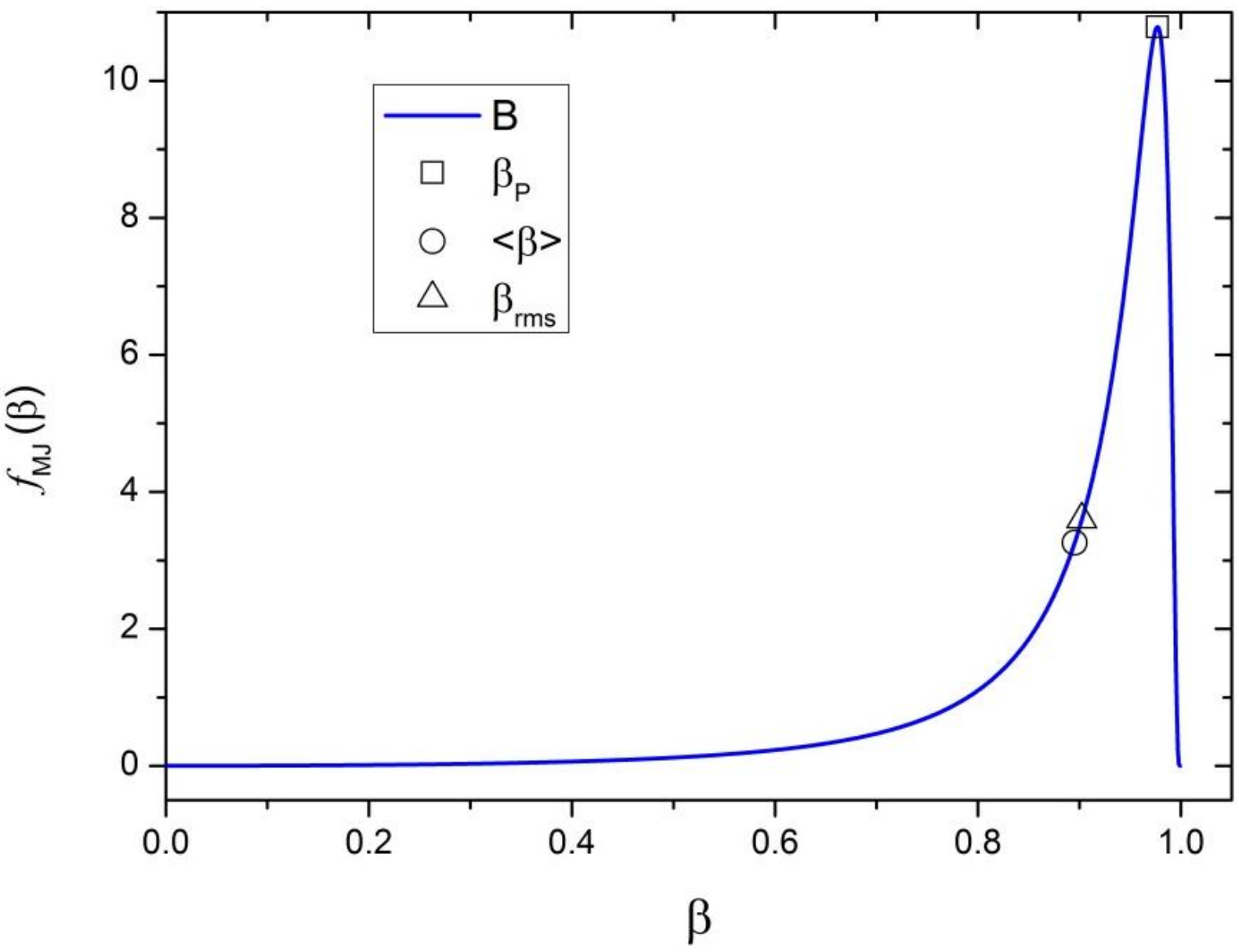

**Figura 8** Distribuição de Maxwell-Jüttner para prótons à temperatura do quasar 3C 273.

Observa-se que a semelhança entre as curvas das distribuições de Maxwell-Boltzmann e Maxwell-Jüttner para $\zeta = 39{,}5335$ traduz-se na proximidade entre os valores esperados e no reduzido erro percentual, o que não ocorre quando o parâmetro $\zeta$ aproxima-se da unidade.

## 5. Comentários Finais

Neste trabalho foi deduzida a expressão da distribuição de velocidades de Maxwell-Jüttner, a qual leva em consideração a energia relativística. Esta distribuição foi comparada com a bem conhecida distribuição de Maxwell-Boltzmann, que não considera efeitos relativísticos. Aplicações e comparações entre estas duas distribuições foram realizadas em três situações: *i*) prótons a temperatura da superfície solar, *ii*) elétrons a temperatura do critério de Lawson para reações D-D, e *iii*) prótons a temperaturas de quasares. O parâmetro fundamental na utilização de uma ou outra distribuição é o parâmetro zeta ($\zeta = m_0c^2/kT$) que leva em conta a massa de repouso da partícula considerada e a temperatura do sistema. Observa-se, pelas Tabelas 2, 4, 5 e 6, que à medida que zeta diminui, a diferença entre as duas distribuições torna-se maior. Zeta é, então, o parâmetro que define quando se deve considerar os efeitos relativísticos e consequentemente a distribuição de velocidades de Maxwell-Jüttner.

A distribuição de velocidades de Maxwell-Boltzmann geralmente é apresentada, sem ser demonstrada, nas disciplinas introdutórias do curso de graduação em Física no tópico sobre teoria cinética dos gases. A sua demonstração costuma ser realizada em disciplinas mais avançadas do curso de graduação, geralmente na disciplina de Física Estatística. Já a distribuição relativística de velocidades de Maxwell-Jüttner, em razão de sua complexidade e especificidade, não faz parte do conteúdo ministrado durante o curso de graduação em Física. Talvez esta pudesse ser inserida em uma disciplina de pós-graduação cujas linhas de pesquisa estejam relacionadas com o referido tópico. Uma sugestão seria o professor em conjunto com os alunos interessados no tema, trabalharem as distribuições de velocidades de Maxwell-Boltzmann e de Maxwell-Jüttner com o software GeoGebra [35]. Isto poderia ser feito de forma mais qualitativa usando o GeoGebra verificando as diferenças de cada uma das distribuições de forma gráfica, sem entrar em muitos detalhes nos extensos cálculos.

Finalizando, enfatizamos que a distribuição de velocidades de Maxwell-Jüttner compartilha algumas limitações das distribuições para o gás ideal clássico: negligência as interações e efeitos quânticos. Uma limitação adicional (não importante no gás ideal clássico) é que a distribuição de Maxwell-Jüttner negligencia as antipartículas.

## Apêndice A. Dedução da constante de normalização $Z_{MJ}$

Para determinar a constante de normalização $Z_{MJ}$ façamos a transformação

$$\tanh(\theta) = \beta, \tag{A.1}$$

e o elemento diferencial $d\beta$ em função de $d\theta$ é

$$d\beta = \frac{d\theta}{\cosh^2(\theta)}. \tag{A.2}$$

Substituindo (A.1) e (A.2) na Eq. (18) tem-se

$$Z_{MJ} = \int_0^{+\infty} e^{-\zeta\cosh(\theta)}\sinh^2(\theta)\cosh(\theta)\, d\theta, \tag{A.3}$$

onde

$$\zeta = \frac{m_0 c^2}{kT} \tag{A.4}$$

é o parâmetro zeta, cuja análise será útil para determinar a necessidade de tratamento relativístico [20]. A mudança nos limites de integração baseia-se na Eq. (A.1), pois, se $\beta$ tende a zero, $\tanh(\theta)$ e $\theta$ também o fazem, ao passo que, se $\beta$ tende a 1, $\tanh(\theta)$ tende a 1, o que resulta na tendência de $\theta$ a infinito. Usando a relação (veja a demonstração no fim do Apêndice A)

$$\sinh^2(\theta)\cosh(\theta) = \left(\frac{1}{4}\right)[\cosh(3\theta) - \cosh(\theta)], \tag{A.5}$$

a Eq. (A.3) fica

$$Z_{MJ} = \frac{1}{4}\left(\int_0^{+\infty} e^{-\zeta\cosh(\theta)}\cosh(3\theta)\, d\theta - \int_0^{+\infty} e^{-\zeta\cosh(\theta)}\cosh(\theta)\, d\theta\right). \tag{A.6}$$

Estas integrais são do tipo

$$K_\nu(\zeta) = \int_0^{+\infty} e^{-\zeta\cosh(t)}\cosh(\nu t)\, dt, \tag{A.7}$$

onde $K_\nu(\zeta)$ denota a função modificada de Bessel de segunda espécie com parâmetros $\nu$ e $\zeta$ [18,19] (veja Apêndice B). Dessa forma, a expressão (A.6) fica

$$Z_{MJ} = \frac{1}{4}[K_3(\zeta) - K_1(\zeta)] \,. \tag{A.8}$$

Tal resultado satisfaz a seguinte relação de recorrência [36]

$$K_{\nu+1}(\zeta) - K_{\nu-1}(\zeta) = \left(\frac{2\nu}{\zeta}\right) K_\nu(\zeta) \,. \tag{A.9}$$

com $\nu = 2$. Assim a expressão (A.8) para $Z_{MJ}$ fica somente

$$Z_{MJ} = \frac{K_2(\zeta)}{\zeta} \,, \tag{A.10}$$

onde no numerador $K_2$ é função modificada de Bessel de segunda espécie e de segunda ordem [18], cujo parâmetro variável é $\zeta$, que, por sua vez, depende da temperatura e da massa das partículas.

A Eq. (A.5) pode ser deduzida da seguinte forma. Seja a relação trigonometria hiperbólica [37]

$$\cosh^2(x) - \sinh^2(x) = 1 \,,$$

Multiplicando esta última expressão por $\cosh(x)$, tem-se

$$\sinh^2(x)\cosh(x) = \cosh^3(x) - \cosh(x) \,. \tag{A.11}$$

Além disso

$$\cosh(2x) = \cosh^2(x) + \sinh^2(x),$$

que multiplicada por $\cosh(x)$ fica

$$\cosh(2x)\cosh(x) = \cosh^3(x) + \sinh^2(x)\cosh(x),$$

$$\cosh(2x)\cosh(x) - \sinh^2(x)\cosh(x) = \cosh^3(x) \,. \tag{A.12}$$

Outra relação trigonométrica útil é

$$\sinh(2x) = 2\sinh(x)\cosh(x),$$

que multiplicada por $\sinh(x)$ fica

$$\sinh(2x)\sinh(x) = 2\sinh^2(x)\cosh(x), \tag{A.13}$$

$$\frac{\sinh(2x)\sinh(x)}{2} = \sinh^2(x)\cosh(x) \,. \tag{A.14}$$

Subtraindo (A.11) de (A.12), tem-se

$$\cosh(2x)\cosh(x) - 2\sinh^2(x)\cosh(x) = \cosh(x),$$

$$2\sinh^2(x)\cosh(x) = \cosh(2x)\cosh(x) - \cosh(x) \,. \tag{A.15}$$

Introduzindo (A.13) em (A.15) tem-se

$$2\left(\frac{\sinh(2x)\sinh(x)}{2}\right) = \cosh(2x)\cosh(x) - \cosh(x),$$

$$\cosh(2x)\cosh(x) - \sinh(2x)\sinh(x) - \cosh(x) = 0\,.$$

Adicionando o termo $2\sinh(2x)\sinh(x)$ em ambos os lados desta última equação

$$\cosh(2x)\cosh(x) + \sinh(2x)\sinh(x) - \cosh(x) = 2\sinh(2x)\sinh(x). \qquad (A.16)$$

Substituindo a seguinte relação

$$\cosh(3x) = \cosh(2x)\cosh(x) + \sinh(2x)\sinh(x),$$

em (A.16), tem-se

$$\cosh(3x) - \cosh(x) = 2\sinh(2x)\sinh(x). \qquad (A.17)$$

Introduzindo (A.13) em (A.17)

$$\cosh(3x) - \cosh(x) = 2[2\sinh^2(x)\cosh(x)],$$

$$4\sinh^2(x)\cosh(x) = \cosh(3x) - \cosh(x),$$

e finalmente

$$\sinh^2(x)\cosh(x) = \left(\frac{1}{4}\right)[\cosh(3x) - \cosh(x)]. \qquad (A.18)$$

## Apêndice B. Função modificada de Bessel

A equação diferencial de Bessel é dada por [18]

$$z^2\frac{d^2w}{dz^2} + z\frac{dw}{dz} + (z^2 - \nu^2)w = 0.$$

A equação diferencial modificada de Bessel de ordem $\nu$ é obtida com a simples alteração de sinal do termo $z^2$ do coeficiente de $w$

$$z^2\frac{d^2w}{dz^2} + z\frac{dw}{dz} - (z^2 + \nu^2)w = 0.$$

As funções modificadas de Bessel são soluções desta equação. Denomina-se $I_\nu(z)$ a função modificada de Bessel de primeira espécie, a saber

$$I_\nu(z) = \sum_{n=0}^{\infty}\frac{(z/2)^{2n+\nu}}{n!\,\Gamma(1+n+\nu)}\,.$$

A função modificada de Bessel de segunda espécie, representada por $K_\nu(z)$, consiste em uma série de potências que se relaciona com $I_\nu(z)$ da seguinte forma

$$K_\nu(z) = \frac{I_{-\nu}(z) - I_\nu(z)}{(2/\pi)\sin(\nu\pi)}\,.$$

Quando $\nu$ é um número $n \in \mathbb{Z}$, toma-se o limite $\nu \rightarrow n$ para que o seno do denominador da equação acima não se anule. No Apêndice A, a Eq. (A.7) representa esta função modificada de Bessel por meio de uma integral.

## Apêndice C. Maxwell-Jüttner para $v/c \ll 1$

A distribuição de Maxwell-Jüttner é dada por

$$f_{MJ}(\beta) = \frac{\zeta\beta^2\gamma^5 e^{-\zeta\gamma}}{K_2(\zeta)} \ ,$$

No limite de baixas velocidades ($v \ll c$ e, consequentemente, $v/c \ll 1$), tem-se

$$K_2(\zeta) \approx \sqrt{\frac{\pi}{2\zeta}} e^{-\zeta},$$

$$(1-\beta^2)^{-1/2} \approx 1 + \frac{\beta^2}{2},$$

$$f_{MJ}(\beta) \approx \zeta\beta^2\gamma^5 e^{-\zeta\gamma} \sqrt{\frac{2\zeta}{\pi}} e^{\zeta},$$

$$f_{MJ}(\beta) \approx \sqrt{\frac{2}{\pi}} \zeta^{3/2}\beta^2 \left(1 + \frac{5}{2}\beta^2\right) e^{-\zeta(1+\beta^2/2)} e^{\zeta},$$

$$f_{MJ}(\beta) \approx \sqrt{2/\pi}\,\zeta^{3/2}\beta^2 \left(1 + \frac{5}{2}\beta^2\right) e^{-\zeta\beta^2/2} \ .$$

Considerando válida a aproximação

$$1 + \frac{5}{2}\beta^2 \approx 1 \, ,$$

tem-se

$$f_{MJ}(\beta) \approx \sqrt{\frac{2}{\pi}} \zeta^{3/2}\beta^2 e^{-\zeta\beta^2/2},$$

que é a distribuição de Maxwell-Boltzmann dada pela Eq. (1), porém, em função de $\zeta$ e $\beta$.

## Apêndice D. Dedução da Eq. (27) da Seção 3

A derivação da distribuição de Maxwell-Jüttner e sua igualdade a zero resulta na Eq. (26)

$$\beta^2\left(1 - \zeta\sqrt{1-\beta^2}\right) - 3\beta^4 + 2 = 0,$$

$$-3\beta^4 + 2 + \beta^2 - \beta^2\zeta\sqrt{1-\beta^2} = 0.$$

A fim de gerar um produto da diferença pela soma e, com isso, tirar todas as raízes quadradas da equação anterior multiplica-se esta última expressão pela seguinte fração

$$\left(-3\beta^4+2+\beta^2-\beta^2\zeta\sqrt{1-\beta^2}\right)\left(\frac{-3\beta^4+2+\beta^2+\beta^2\zeta\sqrt{1-\beta^2}}{-3\beta^4+2+\beta^2+\beta^2\zeta\sqrt{1-\beta^2}}\right)=0\,,$$

$$9\beta^8+(\zeta^2-6)\beta^6-(\zeta^2+11)\beta^4+4\beta^2+4=0.$$

Verifica-se por simples inspeção que tanto 1 quanto $-1$ são raízes desta última equação. Logo, pode-se reduzi-la a um polinômio de grau 6, da seguinte forma

$$(\beta-1)(\beta+1)[9\beta^6+(\zeta^2+3)\beta^4-8\beta^2-4]=0,$$

e como a intenção é obter valores de $\beta$ entre 0 e 1, podemos considerar somente a expressão entre colchetes, ou seja

$$9\beta^6+(\zeta^2+3)\beta^4-8\beta^2-4=0,$$

que é a Eq. (27).

## Apêndice E. Função gama incompleta (Seção 3)

A função gama incompleta é definida pela integral [24]

$$\Gamma(a\,,b)=\int_b^\infty t^{a-1}e^{-t}dt.$$

Na Seção 3, veja Eq. (34), para o cálculo de $\langle\beta\rangle$, tem-se $a=3$, isto é

$$\Gamma(3\,,b)=\int_b^\infty t^2e^{-t}dt.$$

Esta integral possui a solução analítica

$$\Gamma(3\,,b)=\frac{e^{-b}(b^3+2b^2+2b)}{b}.$$

Tomando $b=\zeta/\sqrt{1-\beta^2}$, e aplicando os limites de integração da variável $\beta$ na Eq. (35), tem-se

$$\Gamma\left(3,\frac{\zeta}{\sqrt{1-\beta^2}}\right)\Bigg|_0^1=\lim_{\beta\to1}\Gamma\left(3,\frac{\zeta}{\sqrt{1-\beta^2}}\right)-\Gamma(3,\zeta)\,.$$

Quando $\beta$ tende a 1, $\sqrt{1-\beta^2}$ tende a 0, e $\zeta/\sqrt{1-\beta^2}$ tende a infinito. Em tal situação,

$$\lim_{b\to\infty}\Gamma(3,b)=0\,.$$

Portanto

$$\Gamma\left(3,\frac{\zeta}{\sqrt{1-\beta^2}}\right)\Bigg|_0^1=-\Gamma(3,\zeta)\,.$$

## Apêndice F. Cálculo de $\langle\beta^2\rangle$ (Seção 3)

Seja a integral

$$\langle\beta^2\rangle = \int_0^1 \frac{\zeta\beta^4\gamma^5 e^{-\zeta\gamma}}{K_2(\zeta)} d\beta.$$

Busca-se determinar, como solução, uma expressão que dependa de funções matemáticas especiais, mesmo que sejam dadas por operações com séries de potências. Fazendo a mesma transformação de variável realizada na Seção 3, isto é

$$\sinh(\theta) = \beta\gamma,$$

$$\cosh^2(\theta) - (\beta\gamma)^2 = 1,$$

$$\cosh(\theta) = \sqrt{1 + (\beta\gamma)^2}\ ,$$

$$\cosh(\theta) = \gamma = \frac{1}{\sqrt{1-\beta^2}}\ ,$$

$$\tanh(\theta) = \beta,$$

$$d\beta = \frac{d\theta}{\cosh^2(\theta)}.$$

Expressando $\langle\beta^2\rangle$ em função de $\theta$

$$\langle\beta^2\rangle = \frac{\zeta}{K_2(\zeta)} \int_0^{+\infty} e^{-\zeta\cosh(\theta)} \sinh^3(\theta)\tanh(\theta)\, d\theta,$$

$$\langle\beta^2\rangle = \frac{\zeta}{K_2(\zeta)} \int_0^{+\infty} e^{-\zeta\cosh(\theta)} \sinh^2(\theta)\tanh^2(\theta)\cosh(\theta)\, d\theta.$$

Aplicando a igualdade $\sinh^2(\theta)\tanh^2(\theta) = \sinh^2(\theta) - \tanh^2(\theta)$, tem-se

$$\langle\beta^2\rangle = \frac{\zeta}{K_2(\zeta)} \int_0^{+\infty} e^{-\zeta\cosh(\theta)} \cosh(\theta)[\sinh^2(\theta) - \tanh^2(\theta)]\, d\theta,$$

$$\langle\beta^2\rangle = \frac{\zeta}{K_2(\zeta)} \int_0^{+\infty} \left[e^{-\zeta\cosh(\theta)} \sinh^2(\theta)\cosh(\theta) - e^{-\zeta\cosh(\theta)} \tanh^2(\theta)\cosh(\theta)\right] d\theta.$$

Conforme demonstrado na Seção 3, temos

$$\int_0^{+\infty} e^{-\zeta\cosh(\theta)} \sinh^2(\theta)\cosh(\theta)\, d\theta = \frac{K_2(\zeta)}{\zeta}.$$

Como $\tanh^2(\theta) = 1 - \text{sech}^2(\theta)$, tem-se

$$\langle\beta^2\rangle = \frac{\zeta}{K_2(\zeta)}\left[\frac{K_2(\zeta)}{\zeta} - \int_0^{+\infty} e^{-\zeta\cosh(\theta)}\cosh(\theta)\,d\theta + \int_0^{+\infty} e^{-\zeta\cosh(\theta)}\,\text{sech}(\theta)\,d\theta\right],$$

e usando a relação

$$K_\nu(\zeta) = \int_0^{+\infty} e^{-\zeta cosh(t)}\cosh(\nu t)\,dt,$$

temos

$$\langle\beta^2\rangle = \frac{\zeta}{K_2(\zeta)}\left[\frac{K_2(\zeta)}{\zeta} - K_1(\zeta) + \int_0^{+\infty} e^{-\zeta\cosh(\theta)}\,\text{sech}(\theta)\,d\theta\right].$$

A integral dentro do colchetes desta última equação pode ser resolvida por derivação sob o sinal de integral [38] para uma variável $\alpha$ tal que

$$f(\alpha) = \int_0^{+\infty} e^{-\alpha\zeta\cosh(\theta)}\,\text{sech}(\theta)\,d\theta,$$

$$\frac{d(f(\alpha))}{d\alpha} = f'(\alpha) = -\int_0^{+\infty} \zeta e^{-\alpha\zeta\cosh(\theta)}\cosh(\theta)\,\text{sech}(\theta)\,d\theta,$$

$$f'(\alpha) = -\zeta\int_0^{+\infty} e^{-\alpha\zeta\cosh(\theta)}d\theta,$$

$$f'(\alpha) = -\zeta K_0(\alpha\zeta),$$

$$f(\alpha) = -\zeta\int K_0(\alpha\zeta)d\alpha\,.$$

O resultado desta integral é

$$f(\alpha) = -\frac{\pi\zeta\alpha}{2}[L_{-1}(\zeta\alpha)K_0(\zeta\alpha) + L_0(\zeta\alpha)K_1(\zeta\alpha)] + C,$$

onde $L_\nu(z)$ são funções modificadas de Struve de ordem $\nu$ [25] (veja Apêndice G). Logo,

$$f(\alpha) = \int_0^{+\infty} e^{-\alpha\zeta\cosh(\theta)}\,\text{sech}(\theta)\,d\theta = -\frac{\pi\zeta\alpha}{2}[L_{-1}(\zeta\alpha)K_0(\zeta\alpha) + L_0(\zeta\alpha)K_1(\zeta\alpha)] + C\,.$$

Para determinar a constante $C$, o caminho mais simples seria fazer $\alpha = 0$. Para a expressão

$$f(\alpha) = \int_0^{+\infty} e^{-\alpha\zeta\cosh(\theta)}\,\text{sech}(\theta)\,d\theta\,,$$

tal substituição não acarretaria inconsistências, pois

$$f(0) = \int_{0}^{+\infty} \operatorname{sech}(\theta)\, d\theta,$$

$$f(0) = 2\arctan\left[\tanh\left(\frac{\theta}{2}\right)\right]\Big|_{0}^{+\infty} = \frac{\pi}{2}.$$

No entanto, a expressão

$$f(\alpha) = -\frac{\pi\zeta\alpha}{2}\left[L_{-1}(\zeta\alpha)K_0(\zeta\alpha) + L_0(\zeta\alpha)K_1(\zeta\alpha)\right] + C$$

apresentaria indeterminações, pois ocorreriam os produtos $L_{-1}(\zeta\alpha)K_0(\zeta\alpha)\alpha$ e $L_0(\zeta\alpha)K_1(\zeta\alpha)\alpha$. Considerando que

$$\lim_{\alpha\to 0} K_0(\zeta\alpha) = \infty\,,$$

e

$$K_1(\zeta\alpha) = \tilde{\infty}\,,$$

onde $\tilde{\infty}$ representa o infinito complexo, ter-se-iam duas multiplicações de zero por infinito. Tomando o limite

$$\lim_{\alpha\to 0} f(\alpha) = \lim_{\alpha\to 0}\left\{-\frac{\pi\zeta\alpha}{2}\left[L_{-1}(\zeta\alpha)K_0(\zeta\alpha) + L_0(\zeta\alpha)K_1(\zeta\alpha)\right] + C\right\},$$

têm-se, como resultados, os seguintes limites válidos

$$\lim_{\alpha\to 0} K_0(\zeta\alpha)\,\alpha = 0\,,$$

$$\lim_{\alpha\to 0} L_{-1}(\zeta\alpha)K_0(\zeta\alpha)\,\alpha = \left[\lim_{\alpha\to 0} K_0(\zeta\alpha)\,\alpha\right] L_{-1}(0) = 0[L_{-1}(0)] = 0\,,$$

$$\lim_{\alpha\to 0} L_0(\zeta\alpha)K_1(\zeta\alpha) = \frac{2}{\pi}\,,$$

$$\lim_{\alpha\to 0} L_0(\zeta\alpha)K_1(\zeta\alpha)\alpha = \left[\lim_{\alpha\to 0} L_0(\zeta\alpha)K_1(\zeta\alpha)\right] 0 = 0\,.$$

Portanto

$$\lim_{\alpha\to 0} f(\alpha) = C.$$

Como

$$f(0) = \lim_{\alpha\to 0} f(\alpha) = \lim_{\alpha\to 0} \int_{0}^{+\infty} e^{-\alpha\zeta\cosh(\theta)} \operatorname{sech}(\theta)\, d\theta\,,$$

tem-se que

$$C = \frac{\pi}{2}.$$

Tomando $\alpha = 1$, encontra-se o resultado da integral de interesse

$$f(1) = \int_0^{+\infty} e^{-\zeta \cosh(\theta)} \operatorname{sech}(\theta)\, d\theta = -\frac{\pi\zeta}{2}[L_{-1}(\zeta)K_0(\zeta) + L_0(\zeta)K_1(\zeta)] + \frac{\pi}{2},$$

e tem-se, finalmente, que

$$\langle \beta^2 \rangle = \frac{\zeta}{K_2(\zeta)} \left\{ \frac{K_2(\zeta)}{\zeta} - K_1(\zeta) + \frac{\pi}{2}[1 - \zeta K_0(\zeta)L_{-1}(\zeta) - \zeta K_1(\zeta)L_0(\zeta)] \right\},$$

ou apenas

$$\langle \beta^2 \rangle = 1 - \frac{\zeta K_1(\zeta)}{K_2(\zeta)} + \frac{\zeta\pi}{2K_2(\zeta)}[1 - \zeta K_0(\zeta)L_{-1}(\zeta) - \zeta K_1(\zeta)L_0(\zeta)].$$

## Apêndice G. Função modificada de Struve

A equação diferencial de Struve é dada por [18]

$$z^2 \frac{d^2w}{dz^2} + z\frac{dw}{dz} + (z^2 - \nu^2)w = \frac{4(z/2)^{\nu+1}}{\Gamma(\nu + 1/2)\sqrt{\pi}}.$$

De acordo com [39], trata-se de uma equação diferencial de Bessel não homogênea. A mesma alteração feita no coeficiente de $w$ da equação de Bessel é suficiente, na equação acima, para gerar a equação modificada de Struve, que, conforme [32], se expressa por

$$z^2 \frac{d^2w}{dz^2} + z\frac{dw}{dz} - (z^2 + \nu^2)w = \frac{4(z/2)^{\nu+1}}{\Gamma(\nu + 1/2)\sqrt{\pi}}.$$

A função modificada de Struve $L_\nu(z)$, que ocorre na Eq. (38) da Seção 3, é solução desta equação. Esta função representa a seguinte série de potências

$$L_\nu(z) = \sum_{n=0}^{\infty} \frac{(z/2)^{2n+\nu+1}}{\Gamma(n + \nu + 3/2)\Gamma(n + 3/2)}.$$

## Apêndice H. Distribuição de Maxwell-Boltzmann em função de $\beta$ e $\zeta$ (Seção 4)

A função de distribuição de velocidades de Maxwell-Boltzmann é dada por

$$f_{MB}(v) = 4\pi \left(\frac{m_0}{2\pi kT}\right)^{3/2} v^2 e^{-\frac{m_0 v^2}{2kT}}. \tag{H.1}$$

Para fins de comparação entre a distribuição de Maxwell-Boltzmann e Maxwell-Jüttner, é mais conveniente colocá-las em função da mesma variável. Adota-se $\beta = v/c$ para aferição do percentual das velocidades em relação a velocidade da luz $c$. Para a distribuição de Maxwell-Boltzmann, tem-se $c\beta = v,$ e $cd\beta = dv$. Assim

$$f_{MB}(\beta)d\beta = 4\pi \left(\sqrt{\frac{m_0}{2\pi kT}}\right)^3 \beta^2 c^2 e^{-\frac{m_0 \beta^2 c^2}{2kT}} c d\beta,$$

$$f_{MB}(\beta)d\beta = 4\pi\left(\sqrt{\frac{m_0}{2\pi kT}}\right)^3 \beta^2 c^3 e^{-\frac{m_0c^2\beta^2}{2kT}} d\beta,$$

$$f_{MB}(\beta)d\beta = 4\pi\left(\sqrt{\frac{m_0c^2}{2\pi kT}}\right)^3 \beta^2 e^{-\frac{m_0c^2\beta^2}{2kT}} d\beta,$$

$$f_{MB}(\beta)d\beta = \frac{4\pi}{\left(\sqrt{2\pi}\right)^3}\left(\sqrt{\frac{m_0c^2}{kT}}\right)^3 \beta^2 e^{-\frac{m_0c^2\beta^2}{2kT}} d\beta.$$

Considerando que $\zeta = m_0c^2/kT$, tem-se

$$f_{MB}(\beta)d\beta = \sqrt{\frac{2}{\pi}}\zeta^{3/2}\beta^2 e^{-\zeta\beta^2/2}d\beta$$

ou

$$f_{MB}(\beta) = \sqrt{\frac{2}{\pi}}\zeta^{3/2}\beta^2 e^{-\zeta\beta^2/2},$$

e os valores esperados da distribuição de Maxwell-Boltzmann [21] em função de $\beta$ são

$$\langle\beta\rangle_{MB} = \sqrt{\frac{8kT}{\pi m_0c^2}} = \sqrt{\frac{8}{\pi\zeta}},$$

$$\beta_{p(MB)} = \sqrt{\frac{2kT}{m_0c^2}} = \sqrt{\frac{2}{\zeta}},$$

e

$$\beta_{rms(MB)} = \sqrt{\langle\beta^2\rangle_{MB}} = \sqrt{\frac{3kT}{m_0c^2}} = \sqrt{\frac{3}{\zeta}}.$$

## Referências